\documentclass[conference, 10pt]{IEEEtran}

\usepackage{cite}
\usepackage{amsmath,amssymb,amsfonts}
\usepackage{url}
\usepackage{lipsum}  
\usepackage{bm}
\usepackage{graphicx}
\usepackage{color}
\usepackage{placeins}
\usepackage{float}
\usepackage{soul}
\usepackage{stfloats}
\usepackage{microtype}
\usepackage[caption=false,font=normalsize,labelfont=sf,textfont=sf]{subfig}
\usepackage{tabularx,colortbl}
\usepackage{ifthen}

\makeatletter

\newcounter{author}
\renewcommand{\author}[2][]{
   \stepcounter{author}
   \@namedef{author@\theauthor}{#2}
   \@namedef{authorlabel@\theauthor}{#1}
}

\newcounter{address}
\newcommand{\address}[2][]{
   \stepcounter{address}
   \@namedef{address@\theaddress}{#2}
   \@namedef{addresslabel@\theaddress}{#1}
}

\newcommand{\alsep}{and}

\def\newmaketitle{\par%
  \begingroup%
  \normalfont%
  \def\thefootnote{}
  \def\footnotemark{}
  \let\@makefnmark\relax
  \footnotesize
  \footnotesep 0.7\baselineskip
  \normalsize%
  \twocolumn[\thenewmaketitle\@IEEEaftertitletext]%
  \if@IEEEusingpubid
     \enlargethispage{-\@IEEEpubidpullup}%
  \fi
  \endgroup
  \setcounter{footnote}{0}\let\maketitle\relax\let\@maketitle\relax
  \gdef\@thanks{}%
  \let\thanks\relax}

\def\thenewmaketitle{
  \newpage
  \begin{center}%
    \vskip0.2em{\Huge\@IEEEcompsoconly{\sffamily}\@IEEEcompsocconfonly{\normalfont\normalsize\vskip 2\@IEEEnormalsizeunitybaselineskip
   \bfseries\large}\@title\par}\vskip1.0em\par%
    \vspace{1ex}
    \newcounter{c@author}
    \newcounter{c@tmp}
    \ifthenelse{\value{author}=2}{%
      \newcommand{\liand}{ and }}{%
      \newcommand{\liand}{, and }}
    \ifthenelse{\value{address}<2}{%
      \@nameuse{author@1}%
      \stepcounter{c@author}%
      \whiledo{\value{c@author}<\value{author}}{%
        \setcounter{c@tmp}{\value{author}}%
        \addtocounter{c@tmp}{-\value{c@author}}%
        \ifthenelse{\value{c@tmp}=1}{%
          \renewcommand{\alsep}{\liand}}{\renewcommand{\alsep}{, }}%
        \stepcounter{c@author}\alsep \@nameuse{author@\thec@author}}\\%
    }
    {
      \@nameuse{author@1}${}^{(\ref{\@nameuse{authorlabel@1}})}$%
      \stepcounter{c@author}%
      \whiledo{\value{c@author}<\value{author}}{%
      \setcounter{c@tmp}{\value{author}}%
      \addtocounter{c@tmp}{-\value{c@author}}%
      \ifthenelse{\value{c@tmp}=1}{%
        \renewcommand{\alsep}{\liand}}{\renewcommand{\alsep}{, }}%
      \stepcounter{c@author}\alsep \@nameuse{author@\thec@author}%
        ${}^{(\ref{\@nameuse{authorlabel@\thec@author}})}$%
      }
    }
    \vspace{0.2ex}

    \ifthenelse{\value{address}>0}{%
      \ifthenelse{\value{address}=1}{
        {\@nameuse{address@1}}
      }
      {
        \newcounter{c@address}

        \begin{center}
        \whiledo{\value{c@address}<\value{address}}
        {
          \refstepcounter{c@address}
            ${}^{(\thec@address)}$\,%
              \label{\@nameuse{addresslabel@\thec@address}}%
              \@nameuse{address@\thec@address}\\ %
        }
        \end{center}
      } 
    }
    {
      \relax
    }
  \end{center}
}

\makeatother

\usepackage{eso-pic}
\AddToShipoutPicture*{
\footnotesize\sffamily\raisebox{0.8cm}{\hspace{1.4cm}\fbox{
\parbox{\textwidth}{
© 2024 IEEE.  Personal use of this material is permitted.  Permission from IEEE must be obtained for all other uses, in any current or future media, including reprinting/republishing this material for advertising or promotional purposes, creating new collective works, for resale or redistribution to servers or lists, or reuse of any copyrighted component of this work in other works.}}}}

\title{Block Structure Preserving Model Order Reduction for A-EFIE Integral Equation Method}

\author[org1]{Riccardo Torchio}
\author[org2]{Sebastian Schöps}
\author[org1]{Francesco Lucchini}
\address[org1]{Department of Industrial Engineering, University of Padova, Italy}
\address[org2]{Computational Electromagnetics Group, Technische Universität Darmstadt, Germany}

\begin{document}

\newmaketitle

\begin{abstract}
A Block Structure Preserving Model Order Reduction approach is proposed for Integral Equations methods based on the Augmented Electric Field Integral Equation. This approach allows for representing the unknown fields with dedicated subspaces. Numerical results show that this leads to smaller reduced-order models and higher accuracy. 
\end{abstract}

\section{Introduction}
Integral Equation (IE) methods can be efficiently used to solve large-scale electromagnetic problems. In many applications, the frequency range of interest is large, therefore, IE methods that are stable in the whole frequency range should be used, as the ones based on the discretization of the Augmented Electric Field Integral Equation (A-EFIE), where the EFIE equation is augmented with the continuity equation of the charge density, resulting in a Partial Element Equivalent Circuit (PEEC) like formulation \cite{7423678,4982767,1128204,10455963}.

When frequency sweeps are required, to reduce the overall computation time, Model Order Reduction (MOR) techniques can be used to generate a Reduced Order Model (ROM) that is finally used to perform the frequency sweep simulations efficiently. %
The projection is constructed using a (limited) number of solutions of the Full Order Model (FOM) for some selected frequencies (snapshots). For example, the combination of PEEC and MOR was recently investigated in \cite{9829855,10373844}. Most low frequency-stable formulations are hybrid \cite{481380}, i.e., consist of two or even more unknown fields. This is necessary to separate the electric and magnetic phenomena in the static limit. For example, the A-EFIE is based on currents $\mathbf{j}$ and potentials $\bm{\phi}$ or charges $\mathbf{q}$ as degrees of freedom. However, in the literature, when MOR techniques are used, the projection matrix %
is usually constructed by employing a monolithic strategy, where the whole solution %
is used to construct the projectors. %
However, the drawbacks are well-known, for example in the circuit and thermal modelling communities, see e.g. \cite{1518178,7458469}.

In this paper, we propose the application of the block-structure preserving MOR technique from \cite{1518178}, where $\mathbf{j}$ and $\bm{\phi}$ are projected separately. Numerical results show that this block approach leads to a smaller and more accurate ROM.

\section{A-EFIE Formulation}
The electromagnetic (EM) problem, in the frequency domain, is represented  by the EFIE coupled with the continuity equation of electric current density $\mathbf{J}$, expressed as:
\begin{equation} \label{eq.EFIE}
    \rho(\mathbf{r})\mathbf{J}(\mathbf{r})=-i\omega \mathbf{A}(\mathbf{r})-\nabla \varphi(\mathbf{r}) +\mathbf{E}_{ext}(\mathbf{r})    
\end{equation}
\begin{equation} \label{eq.contJ}
    \nabla\cdot\mathbf{J}(\mathbf{r})=-i\omega\varrho(\mathbf{r}),
\end{equation}
where $\omega=2\pi f=kc_0$ is the angular frequency, $\varrho$ is the charge density, $\mathbf{A}$ and $\varphi$ are the vector and scalar potentials related to electric current and charge densities through integral expressions involving Green's kernel, $\rho$ is the resistivity of the material and $\mathbf{E}_{ext}(\mathbf{r})$ is the external applied electric field. By applying the Galerkin testing procedure to \eqref{eq.EFIE}, the following system of equations can be written for the unknown currents $\mathbf{j}$ and potentials $\bm{\phi}$:
\begin{equation}\label{eq.sys}
    \underbrace{\begin{bmatrix}
    \mathbf{R} + i\omega\mathbf{L} & \mathbf{S}^\top \\
    \mathbf{P}\mathbf{S} & -i\omega\mathbf{1}
    \end{bmatrix}}_{\textbf{A}}
     \underbrace{\begin{bmatrix}
    \mathbf{j}\\
    \bm{\phi}
    \end{bmatrix}}_{\textbf{x}}
    =
    \underbrace{\begin{bmatrix}
    \mathbf{e_{ext}}\\
    \mathbf{0}
    \end{bmatrix}}_{\textbf{b}},
\end{equation}
where $\mathbf{R}$ is the resistance matrix, $\mathbf{L}$ is the inductance matrix, $\mathbf{P}$ is the potential matrix and $\mathbf{S}$ is the discrete counterpart of divergence operator, see \cite{9987653} for more details. 

\section{Block Structure Preserving MOR}
We follow the classical reduced basis approximation methodology with a greedy sampling strategy \cite{Patera_2007aa}. Let us assume that we are interested in the angular frequencies $\omega_1,\omega_2,\ldots,\omega_n$. We first solve \eqref{eq.sys} for a subset $\mathcal{I}\subset\{1,\ldots,n\}$ of frequencies. We collect the corresponding solutions $\mathbf{x}_i$ with $i\in\mathcal{I}$ either in a single matrix $\mathbf{X}$ (monolithic approach) or currents and voltages separately in $\mathbf{X}_1$ and $\mathbf{X}_2$ (block approach). We perform a (two) Gram-Schmidt-procedure(s) to orthonormalize the modes and store them in the projection matrix $\mathbf{V}$ (monolithic) or $\mathbf{V}_1$ and $\mathbf{V}_2$ (block approach). Then, we project the system, either monolithically, i.e. we solve $\tilde{\mathbf{A}}\tilde{\mathbf{x}}=\tilde{\mathbf{b}}$ with 
$\tilde{\mathbf{A}}=\mathbf{V}^\top\mathbf{A}\mathbf{V}$ and $\tilde{\mathbf{b}}=\mathbf{V}^\top\mathbf{b}$, or with the block-preserving approach:
\begin{equation*}%
    \underbrace{\begin{bmatrix}
    \mathbf{V}_1^\top \mathbf{R} \mathbf{V}_1
    + 
    i\omega\mathbf{V}_1^\top\mathbf{L} \mathbf{V}_1
    &
    \mathbf{V}_1^\top\mathbf{S}^\top\mathbf{V}_2 \\
    \mathbf{V}_2^\top\mathbf{P}\mathbf{S}\mathbf{V}_1
    &
    -i\omega\mathbf{V}_2^\top\mathbf{1}\mathbf{V}_2
    \end{bmatrix}}_{\tilde{\mathbf{A}}}
     \underbrace{\begin{bmatrix}
    \tilde{\mathbf{j}}\\
    \tilde{\bm{\phi}}
    \end{bmatrix}}_{\tilde{\textbf{x}}}
    =
    \underbrace{\begin{bmatrix}
    \textbf{V}_1^\top\mathbf{e_{ext}}\\
    \textbf{V}_2^\top\mathbf{0}
    \end{bmatrix}}_{\tilde{\textbf{b}}}.
\end{equation*}
For the remaining frequency points, the reconstructed solution, e.g. $\mathbf{x}_i=\textbf{V}^\top\tilde{\textbf{x}}_i$ from the reduced system is plugged into the FOM \eqref{eq.sys}. If the residual error is below some tolerance, we accept the solution, otherwise, we compute a new high-fidelity solution of the FOM and augment the projection matrices, i.e., $\textbf{V}$ or $\textbf{V}_1$ and $\textbf{V}_2$, after orthonormalization with Gram-Schmidt.

\section{Numerical Results}

\begin{figure}[t]
    \centering
\includegraphics[width=0.45\textwidth]{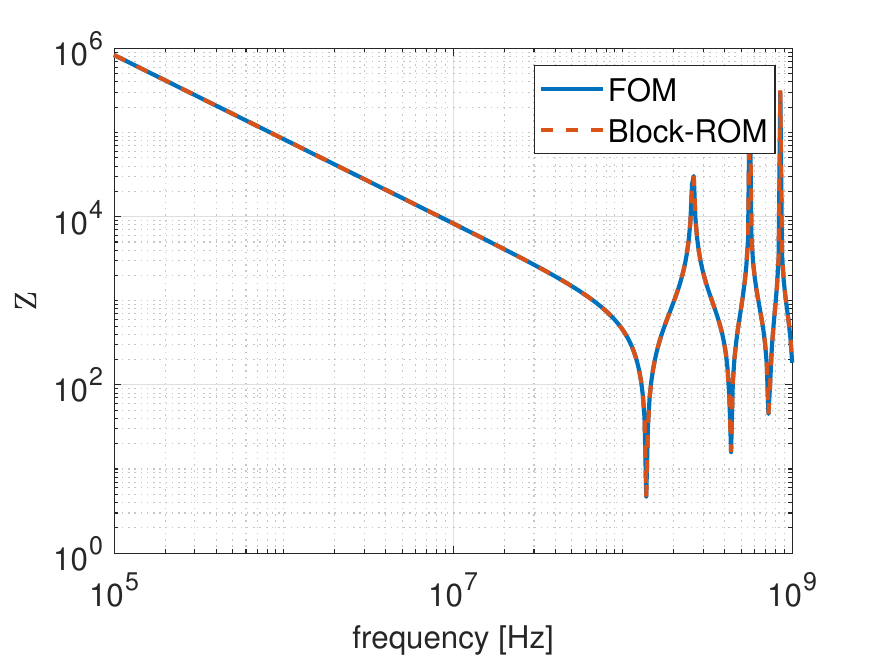}
    \caption{Impedance $Z$ of the antenna over frequency.}
    \label{fig:Z}
\end{figure}
A simple 1~m long electric dipole antenna is considered. The antenna is simulated in a  large frequency range, from 0.1~Hz to 1~GHz. The magnitude of the equivalent impedance Z of the antenna is shown in Fig.~\ref{fig:Z} for a selection of the whole frequency range. 
\begin{figure}[t]
    \centering
\includegraphics[width=0.45\textwidth]{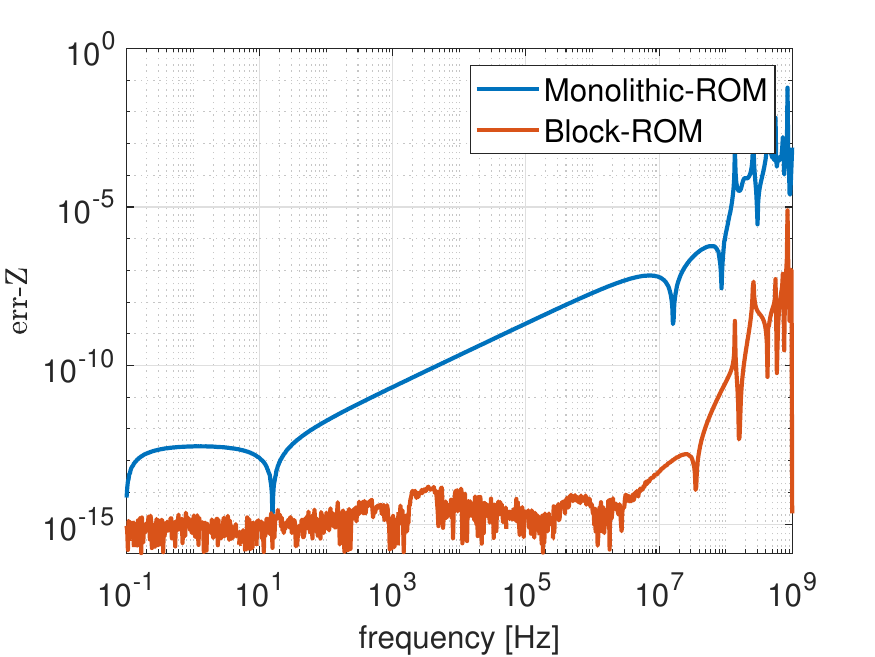}
    \caption{Relative error in the impedance over frequency.}
    \label{fig:errZ}
\end{figure}
Two ROMs have been generated: one using the standard monolithic strategy (i.e., Monolithic-ROM) and the other using the block approach (i.e., Block-ROM). For an accuracy of $10^{-3}$ the monolithic approach needs 41 FOM solutions while the block approach only needs 14 solutions. The size of the FOM is 1000. For some models, the monolithic does not even converge for the target accuracy. Fig.~\ref{fig:errZ}, shows the relative error of the equivalent impedance obtained from the two ROMs w.r.t. the FOM one, i.e., 
\begin{equation}    %
\mathrm{err}\text{-}\mathrm{Z}=|Z_\mathrm{ROM}-Z_\mathrm{FOM}|\slash |Z_\mathrm{FOM}|.
\end{equation}

\begin{figure}[t]
    \centering
\includegraphics[width=0.45\textwidth]{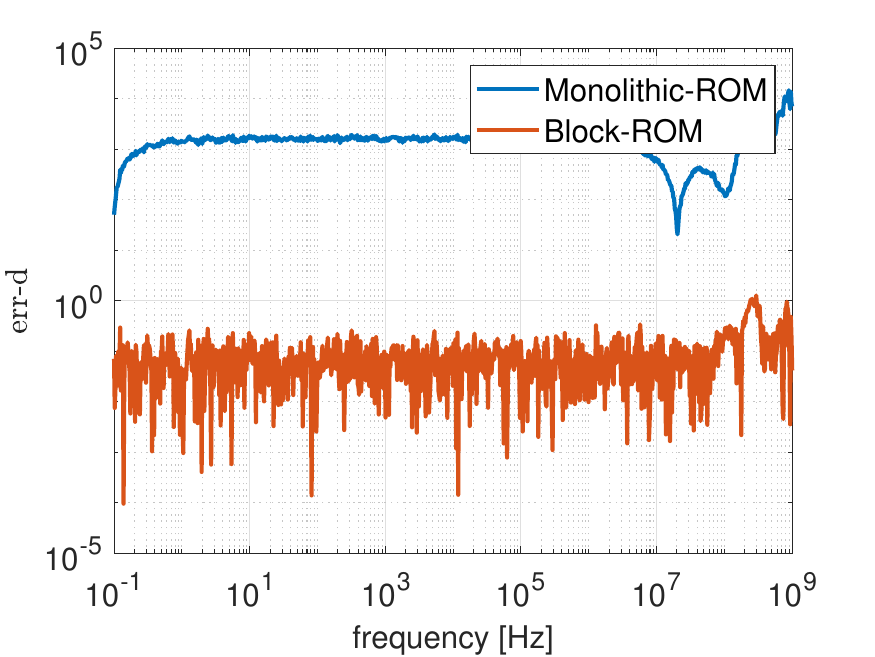}
    \caption{Relative deviation from solenoidality over frequency.}
    \label{fig.PdivJ_phi}
\end{figure}

To show that the monolithic approach is not able to preserve the structure of the problem, the deviation from solenoidality $\mathbf{d}=\mathbf{P}\mathbf{S}^\top\mathbf{j}-i\omega\bm{\phi}\overset{!}{=}0$ is computed, i.e., 
\begin{equation}
\mathrm{err}\text{-}\mathrm{d}=||\mathbf{d}_\mathrm{ROM}-\mathbf{d}_\mathrm{FOM}||\slash ||\mathbf{d}_\mathrm{FOM}||.
\end{equation}
Fig.~\ref{fig.PdivJ_phi} shows the results. The error of the monolithic approach is significantly higher, independent of frequency. 
\section{Conclusions}

A block-structure preserving model order reduction approach has been applied to produce a small and accurate model of electromagnetic devices for a wide frequency range of applications. The approach is applied to an integral equations method based on the augmented electric field integral equation, resulting in a partial-element-equivalent-circuit-like formulation.


\end{document}